
\magnification 1200
\hsize=16 true cm
\vsize=24 true cm
\baselineskip=14pt plus 2pt
\parskip=\baselineskip
\parindent=2.5em

\font\bigbold=cmbx10 scaled\magstep1
\font\medbold=cmbx10 scaled\magstep1

\font\slant=cmsl10

\def\page{\vfill\eject}
\def\title#1{\centerline{}\vskip 1 true cm\centerline{\bigbold #1}}
\def\titleplus#1{\centerline{\bigbold #1}}
\def\authorsinst#1#2{\bigskip\centerline{\medbold #1}\medskip #2\bigskip}
\def\deptiec{{\slant\centerline{
  Departament d'Astronomia i Meteorologia, Universitat de Barcelona, and
	}\centerline{
  Laboratori d'Astrof\'\i sica, Societat Catalana de F\'\i sica,
	}\centerline{
  Institut d'Estudis Catalans.
	}\centerline{
  Avda. Diagonal 647, E-08028 Barcelona, Spain.
	}}}
\def\tobepublished#1{\vskip 3 true cm\centerline{To be published in #1}\page}
\def\headings#1{\par\noindent{\it Subject headings:\/}\ #1}
\def\sec#1{\vskip2\baselineskip\centerline{\bf #1}}
\def\refhead{\vskip\baselineskip\parindent=0pt\par\centerline{\bf
   References}\vskip 0.8\baselineskip\parskip=0pt}
\def\normal{\parskip=\baselineskip\parindent=2.5em}
\def\ref#1{\hangindent=3em\hangafter=1 #1\par}

\def\ie{\it i.e.\/\rm,\ }
\def\eg{\it e.g.\/\rm,\ }

\def\lta{\mathrel{\spose{\lower 3pt\hbox{$\mathchar"218$}}
  \raise 2.0pt\hbox{$\mathchar"13C$}}}
\def\gta{\mathrel{\spose{\lower 3pt\hbox{$\mathchar"218$}}
  \raise 2.0pt\hbox{$\mathchar"13E$}}}
\def\spose#1{\hbox to 0pt{#1\hss}}

\title{TIDALLY INDUCED ELONGATION AND ALIGNMENTS}
\titleplus{OF GALAXY CLUSTERS}
\authorsinst{Eduard Salvador-Sol\'e{\hskip 1em} and{\hskip 1em} Jos\'e
  Mar\'\i a Solanes}{\deptiec}

\sec{ABSTRACT}

We show that tidal interaction among galaxy clusters can account for their
observed alignments and very marked elongation and, consequently, that these
characteristics of clusters are actually consistent with them being formed in
hierarchical clustering. The well-established distribution of projected axial
ratios of clusters with richness class $R\ge 0$ is recovered very
satisfactorily by means of a simple model with no free parameters. The main
perturbers are relatively rich ($R\ge 1$) single clusters and/or groups of
clusters (superclusters) of a wider richness class ($R\ge 0$) located within
a distance of about 65 $h^{-1}$ Mpc from the perturbed cluster. This makes
the proposed scheme be also consistent with all reported alignment effects
involving clusters. We find that this tidal interaction is typically in the
saturate regime (\ie the maximum elongation allowed for systems in
equilibrium is reached), which explains the very similar intrinsic axial
ratio shown by all clusters. Tides would therefore play an important role in
the dynamics of large scale structures, in particular, they should be taken
into account when estimating the virial mass of clusters.

\headings{galaxies: clustering -- celestial mechanics, stellar dynamics --
cosmology: theory}

\tobepublished{The Astrophysical Journal}

\sec{1. INTRODUCTION}
Analyses of large samples of rich galaxy clusters show that their projected
galaxy distribution is on the average considerably elongated (\eg McGillivray
et al. 1976; Carter \& Metcalfe 1980; Binggelli 1982; Di Fazio \& Flin 1988;
Plionis, Barrow \& Frenk 1991), even more than elliptical galaxies (Binggeli
1982). The elongated shape of clusters does not seem to be supported by
rotation (Rood et al. 1972; Gregory \& Tifft 1976; Dressler 1981). The
distribution of observed (projected in 2D) elongations is consistent with
clusters being prolate spheroids (Di Fazio \& Flin 1988; Plionis et al. 1991)
with intrinsic (in 3D) axial ratio Gaussian-distributed with mean $\sim 0.5$,
and standard deviation $\sim 0.15$. It is certainly not consistent with pure
oblate spheroids, although triaxial or combined configurations cannot be
discarded. This result is, in principle, hard to understand in a hierarchical
clustering scenario. Indeed, high density peaks in the primordial density
field are triaxial, and collapse is achieved along the short axis leading to
oblate-like shapes (Peacock \& Heavens 1985; Bardeen et al. 1986).

Prolate spheroids can develop in relaxed spherical systems via bar
instability. However, there is so far no observational evidence of such
highly elongated velocity tensors in galaxy clusters as to favor this kind of
instability. Binggeli's (1982) finding that cluster projected major axis
tends to point towards the {\it nearest} neighboring cluster if its
separation is less than $\sim 15\; h^{-1}$ Mpc seemed to suggest the action
of tidal forces. However, this alignment effect has not been confirmed; it is
found either marginally significant (Flin 1987; Rhee \& Katgert 1987) or
simply insignificant (Struble \& Peebles 1985; Ulmer, McMillan, \& Kowalski
1989; Fong, Stevenson, \& Shanks 1990). Instead, some alignment seems to
exist between the cluster major axis orientation and the direction of {\it
any} cluster neighbor within $30\; h^{-1}$ Mpc, and possibly as much as 50-60
$h^{-1}$ Mpc (Binggeli 1982; West 1989b). Other alignment effects involving
clusters concern the excess of galaxies in the vicinity of these structures
in the direction of their major axis (Argyres et al. 1986; Lambas, Groth, \&
Peebles 1988), and the alignment of galaxy groups with their more or less
elongated spatial distribution in superclusters (West 1989a). All these
alignments would rather point to an intrinsic origin of cluster elongation
related to the aspherical shape of large scale structures. This is naturally
expected from pancake cosmogonies (Oort 1983), an idea supported by N-body
simulations of galaxy formation (Frenk, White, \& Davis 1983; Dekel, West, \&
Aarseth 1984). However, other kinds of observations seem to favor
hierarchical cosmogonies. Besides, there is the clear trend for first ranked
galaxies in clusters to show the same orientation as their parent structures
(Sastry 1968; Carter \& Metcalfe 1980; Binggeli 1982), which, given the
mobility of galaxies and the extreme fragility of their orientation through
merging and capture, is hard to understand in terms of a mere innate effect.
So the possibility that cluster elongation is, after all, tidally induced
should be investigated in detail, particularly if clusters are really prolate
spheroids that form via hierarchical clustering.

Binney \& Silk (1981) have shown that tidal interaction in the linear regime
of density fluctuations should yield prolate shapes with an axial ratio of
protoclusters of $\sim 0.5$, \ie exactly the typical value found in clusters.
This might explain the elongation of not yet collapsed large scale structures
as cluster haloes. However, it can hardly account for the elongation of
clusters themselves because these have evolved through a highly non-linear
phase including violent relaxation, during which any preexisting elongation
is severely damped out (Aarseth \& Binney 1978). A rough estimate of the
typical axial ratio that should prevail after relaxation leads to values of
about $0.7-0.8$ (Binney \& Silk 1981), \ie much higher (or, equivalently, the
shape much less elongated) than observed. However, this estimate does not
take into account that tidal interaction keeps going on after cluster
virialization, which should make the actual elongation be greater.

In the present paper we investigate the elongation induced by tidal
interaction among single and grouped virialized clusters embedded in
non-steady halos. In \S\ 2 we derive the main equations dealing with such an
interaction. In \S\ 3 we determine the distribution of projected axial ratios
and compare it with observation. The expected alignment effects and other
consequences of the proposed scenario are discussed in \S\ 4. \vskip
\baselineskip

\sec{2. TIDALLY-INDUCED ELONGATION}

Let us consider two particle systems, hereafter referred to as perturbed and
perturbing, of masses $M$ and $M'$, respectively, with barycenters separated
by a distance $s$. Let the perturbed system be in steady state (in
particular, with no angular momentum), and its velocity tensor isotropic as a
consequence of violent relaxation (Lynden-Bell 1967). The latter assumption
guarantees sphericity in the lack of appreciable tidal forces, although this
is not mandatory. Indeed, the tidally-induced elongation prior to
virialization might yield some anisotropy in the velocity tensor able to
support by its own some degree of asphericity. However, this anisotropy is
hard to determine. In this sense, the isotropic assumption is a necessary
approximation, with the advantage of notably simplifying the calculations.

In cartesian coordinates, with origin at the barycenter of the perturbed
system and $x_1\/$-axis towards the perturbing one, the integral over the
perturbed system of the $i$ component of the equation of motion of a fluid
element in the global potential field times $x_j$ leads to (Chandrasekhar \&
Lebovitz 1963; Chandrasekhar 1969)
$$
-{2\over 3}\,\tau\; \delta_{ij}=U_{ij}+2\mu\,\delta_{i1}\,
I_{1j}-\mu\,\delta_{i2}\, I_{2j}-\mu\,\delta_{i3}\, I_{3j}.
\eqno(1)
$$
In equation (1) $\tau$ is the
internal kinetic energy, and $U_{ij}$ and $I_{ij}$ are the potential energy
and inertia tensors, respectively, the former defined as
$$
U_{ij}\;=\;
\int{\rho\; {\partial \Phi\over\partial x_i}\;x_j\;d{\hbox{\bf x}}},
\eqno(2)
$$
with $\rho$ the density and $\Phi$ the potential of the perturbed
system alone, and $\mu$ is $$ \mu = {G M' \over s^3}. \eqno(3) $$ In deriving
relation (1) the potential due to the perturbing system has been approximated
by that of a point mass $M'$ located at its barycenter. Since the external
potential of a system is more spherical than the mass distribution causing
it, this should be a good approximation even for relatively small separations
$s$. Besides, it neglects third order terms in $l / s$, with $l$ the size of
the perturbed system, which would give an error of 13 \% in the most
unfavorable case of two interacting systems in physical contact.

For {\it homogeneous} ellipsoids one has (Chandrasekhar 1969)
$$
M\; =\; {4\over3}\;\pi \;\rho\; a_1\; a_2\; a_3,
\eqno(4)
$$
$$
I_{ij}\;=\; {1\over5}\; M\; a_i^2\; \delta_{ij},
\eqno(5)
$$
$$
U_{ij}\;=\; -2\pi\; G\; \rho\; A_i\; I_{ij},
\eqno(6)
$$
with $a_i$ the semi-axis lengths, and some $A_i$ geometrical parameters,
which, for prolate spheroids ($a_1 > a_2 = a_3$), write
$$
A_1\;=\; {2\beta^2\over (1-\beta^2)^{3/2}}\; \biggl[ln \Bigl({\beta\over
1-(1-\beta^2)^{1/2}}\Bigr)\;-\;(1-\beta^2)^{1/2}\biggr],
\eqno(7a)
$$
$$
A_2\;=\;A_3\;=\; {\beta^2\over
(1-\beta^2)^{3/2}}\;\biggl[{(1-\beta^2)^{1/2}\over \beta^2}\;-\;ln
\Bigl({\beta\over 1-(1-\beta^2)^{1/2}}\Bigr)\biggr],
\eqno(7b)
$$
in terms of the {\it axial ratio}
$$
\beta\; =\; {a_2\over a_1}.
\eqno(8)
$$

Taking into account relations (4)--(7$b$), the non-vanishing components of
the tensorial equation (1),
$$
-{2\over3}\,\tau =U_{11}+2\mu\, I_{11}= U_{22}\,-\mu\, I_{22}=U_{33}-\mu\,
I_{33},
\eqno(9)
$$
lead to the relation (Chandrasekhar 1969)
$$
{2\beta^2\over (1-\beta^2)^{3/2}}\; \biggl[ln \Bigl({\beta\over
1-(1-\beta^2)^{1/2}}\Bigr)\;-\;{3\;(1-\beta^2)^{1/2}\over
2+\beta^2}\biggr]\;=\;{\mu\over\pi\; G\; \rho},
\eqno(10)
$$
between the intrinsic axial ratio $\beta$ of the perturbed system and the
mass and distance, through $\mu$, of the perturbing one. Whereas, taking into
account the same relations (4)--(7$b$) and equation (9), the trace of
tensorial equation (1) leads to the virial theorem {\it generalized for the
case of non-negligible tidal elongations}.

The solution $\beta(\mu)$ of equation (10) is bivaluate, each pair of values
corresponding to one stable and one unstable figure of equilibrium. The two
sets of solutions form two different branches of the so-called Jeans
sequence, separated at $\displaystyle {\mu\over \pi\; G\; \rho}= 0.1255$ by
the degenerate solution $\beta_{min}=0.4693$ giving the minimum stable
solution and maximum unstable one. Since we are concerned here only with the
stable branch of the Jeans sequence, we can regard expression (10) as a
unimodal relation between $\mu$ and $\beta$. For values of $\displaystyle
{\mu\over\pi\; G\; \rho}$ larger than 0.1255, there is no solution. Note the
coincidence between the value $\beta_{min}$ and the typical axial ratio of
clusters.

Galaxy clusters are not homogeneous ellipsoids but self-similar inhomogeneous
ones (Salvador-Sol\'e, Sanrom\`a, \& Gonz\'alez-Casado 1992). Under these
circumstances the density distribution can be expressed as a radial density
profile in terms of the equivalent radius $\displaystyle {r=\Bigl({3V\over
{4\pi}}\Bigr)^{1/3}}$, with $V$ the volume inside each homologous isodensity
contour, the total equivalent radius $R$ being equal to the geometrical mean
of the semi-axes $a_1$, $a_2$, and $a_3$. This leads to (Roberts 1962)
$$
M\;=\; {4\pi\over 3}\; \rho(0)\; a_1\;a_2\;a_3\; f_M,
\eqno(11)
$$
$$
I_{ij}\;=\; {1\over 5}\; M\; a_i^2 \; \delta_{ij} \; f_I,
\eqno(12)
$$
$$
U_{ij}\;=\; -2\pi\; G\; \rho(0)\; A_i\; I_{ij} \; f_U,
\eqno(13)
$$
with $A_i$ the geometrical coefficients given by expressions (7$a$) and
(7$b$). So the only difference with the homogeneous case (eqs. \lbrack
4\rbrack --\lbrack 6\rbrack) comes from the shape factors
$$
f_M\; =\; 3 \int_0^1{\eta(x)\; x^2\; dx},
\eqno(14a)
$$
$$
f_I\; =\; {15\over 2\;f_M}\; \int_0^1{F(x)\; x^2\; dx},
\eqno(14b)
$$
$$
f_U\; =\; {15\over 8\;f_I\; f_M}\; \int_0^1{F^2(x)\; dx},
\eqno(14c)
$$
with $\eta(x) = \rho (x)/\rho(0)$, $x=r/R$, and
$$
F(\xi)\;=\; 2 \int_{\xi}^1{\eta(x) \;x \;dx}.
\eqno(14d)
$$
Thus, by substituting $U_{ij}$ and $I_{ij}$ given by relations (12) and (13)
into equation (9) and taking into account relations (11), (7$a$), (7$b$), we
obtain the wanted relation between the axial ratio of the perturbed system,
and the mass and distance of the perturbing one for the case of self-similar
inhomogeneous systems. This relation turns out to be identical to relation
(10) for the homogeneous case, but for the fact that constant density $\rho$
must be replaced by the effective density $\widetilde\rho=\rho(0)\; f_U$.

Now let us consider the case of tidal interaction between a couple of relaxed
clusters with similar shape of their mass density profile, assumed with the
modified Hubble law form, $\eta(x)= \Bigl(1+(xR/r_0)^2\Bigr)^{-\alpha}$, but
different central densities, axial ratios, and orientations. Then equation
(10) with effective density $\widetilde\rho$ can be written explicitly as
$$
G(\beta)\;=\; Q\; {4\; f_M(R/r_0,\alpha) \over 3\; f_U(R/r_0,\alpha)}\;
\Bigl({R\over s}\Bigr)^3,
\eqno(15)
$$
where $G(\beta)$ stands for the left-hand-side member of equation (10) and
$Q$ is the ratio of central densities of the perturbing and the perturbed
clusters. Actually, equation (15) is not yet well-suited for our purposes.
Firstly, galaxy clusters are not isolated steady bodies, but {\it are
surrounded by large non-steady halos}. Secondly, we are not concerned with
the elongation of the whole perturbed system, but {\it only of its galactic
component}.

Non-steady halos are elongated in the same direction as their respective
clusters (Argyres et al. 1986; Lambas, et al. 1988). This is consistent with
the elongation of both subsystems being tidally induced by the same
perturber. Since cluster halos are in the linear or moderately non-linear
regime, their typical axial ratio should be about $0.5$, hence, very similar
to that found in relaxed clusters. Under these circumstances the potential
inside the perturbed system keeps with the same form as above. So by
integrating the {\it i} component of the equation of motion of a fluid
element times $x_j$ over the {\it the steady region alone} and neglecting the
pressure at the edge of that region (as usually done for estimating the
cluster virial mass) we are once again led to equation (1) and, consequently,
to equation (15) but for two small differences: 1) due to the non-vanishing
potential of the {\it perturbed} halo, $f_U(R/r_0,\alpha)$ turns out to be
replaced by
$$
{\widetilde f}_U(R/r_0,\alpha,R_h/r_0,\alpha_h)\;=\;f_U(R/r_0,\alpha)\;
$$
$$
\times \Bigg[1\,+\,{\displaystyle \biggl({R_h\over
R}\biggr)^{2(1-\alpha_h)}-1\over {2-2\alpha_h\over 5}\,\displaystyle
\biggl({R\over r_0}\biggr)^{-2\alpha}\,
f_I(R/r_0,\alpha)\,f_U(R/r_0,\alpha)}\Biggr],
\eqno(16a)
$$
and 2) due to the potential of the {\it perturbing} halo,
$f_M(R/r_0,\alpha)$ is also replaced by
$$
{\widetilde f}_M(R/r_0,\alpha,R_h/r_0,\alpha_h)\;=\;f_M(R/r_0,\alpha)\;
\Biggl[1\;+\;{\displaystyle\biggl({R_h\over R}\biggr)^{3-2\alpha_h}-1\over
(3-2\alpha_h)\;\displaystyle \biggl({R\over
r_0}\biggr)^{2\alpha}\;f_M(R/r_0,\alpha)}\Biggr].
\eqno(16b)
$$
Thus, equation (15) transforms into
$$
G(\beta)\;=\; Q\; {4\;\widetilde f_M(R/r_0,\alpha,R_h/r_0,\alpha_h) \over 3\;
{\widetilde f}_U(R/r_0,\alpha,R_h/r_0,\alpha_h)}\; \Bigl({R\over s}\Bigr)^3.
\eqno(17)
$$
In equations (16$a$) and (16$b$) halos are assumed with total radius $R_h$
and mass density profile, {\it corrected from the uniform mean background
density}, of the power law form with index $-2\alpha_h$ (different from 2 and
3), and amplitude matching the interior mass density run. In calculating the
potentials of the whole perturbing system and of the halo of the perturbed
system one should correct, indeed, their respective density profiles for the
uniform mean density because the potential at any point of the inner
perturbed cluster due to an outer uniform mass distribution vanishes. Since
performing the same correction to the density profile of the inner perturbed
cluster makes a negligible difference because of the very high density
contrast there, we can use  corrected profiles everywhere, which notably
simplifies the modelling.

To deal with the second problem we have followed the same derivation above
from the equation of motion of {\it an element of the galactic component}.
This leads to a tensorial equation similar to (1) but with the inertia tensor
$I^g_{ij}$ written in terms of the mass density of the galactic component
instead of the total mass density, and the potential energy tensor $U^g_{ij}$
in terms of both densities ($\mu$ is always written in terms of the {\it
total} mass of the perturber). It can be shown (see Appendix B) that this
latter tensor has the same form as $U_{ij}$ (eq. \lbrack 13\rbrack), with
$I^g_{ij}$ instead of $I_{ij}$ and the shape factor $f^g_U$ given by equation
(B4). The density profile of the galactic component in the inner relaxed
cluster is also assumed with the modified Hubble law form with identical
total and core radii as for the total mass distribution (final results are
very insensitive to the exact value of the core radius), but a distinct power
index $\alpha^g$. The non-vanishing components of tensorial equation (1) then
lead to an analogous relation to (15)
$$
G(\beta)\,+4\,\bigg({1\over 2+(\beta^g)^2}\;-\;{1\over 2+\beta^2}\biggr)\;=
   \; Q\; {4\; f_M(R/r_0,\alpha) \over 3\; f_U^g(\alpha^g,R/r_0,\alpha)}\;
\Bigl({R\over
   s}\Bigr)^3.
\eqno(18)
$$
Equation (18) tells us that for different shaped density profiles of the
galactic component and the total mass, the axial ratio $\beta^g$ of the
galactic distribution will be different from that of the whole mass
distribution. Of course, as for relation (15), relation (18) presumes the
whole system is in steady state. When non-steady cluster haloes are taken
into account one must replace, as in the analogous relation for the total
mass (eq. \lbrack 17\rbrack), functions $f_M$ and $f_U^g$ by $\widetilde f_M$
and $\widetilde f_U^g$, respectively, with $\widetilde f_U^g$ defined in
terms of $f_U^g$ exactly as ${\widetilde f}_U$ in terms of $f_U$ (eq. \lbrack
16a\rbrack). And by substituting $G\beta)$ by its explicit expression (17),
the new and rather complicated relation can be put in the practical form
$$
{4\over G(\beta)}\;\biggl({1\over 2+(\beta^g)^2}\;-\;{1\over
2+\beta^2}\biggr) \;=\;{{\widetilde f}_U(R/r_0,\alpha,R_h/r_0,\alpha_h)\over
{\widetilde f}_U^g(\alpha^g,R/r_0,\alpha,R_h/r_0,\alpha_h)}\;-\;1,
\eqno(19)
$$
which will be used in next section in order to obtain the observable
distribution of $\beta^g$'s from the much easier to calculate (though hidden
to observation) distribution of $\beta$'s. Note that, for $\beta=1$,
$G(\beta)$ is equal to zero and, therefore, that $\beta^g$ is also unity,
whereas the value of $\beta^g$ in the saturate regime, $\beta^g_{min}$,
differs from $\beta_{min}$.

A last comment is in order before inferring the distribution of axial ratios.
As mentioned above, equation (10) and, consequently, equations (17) and (19)
may have no solution, \ie there may be no figure of equilibrium, for too
strong a tidal interaction or, equivalently, too large an $M'$ and/or small
$s$. But ``relaxed'' clusters are necessarily in equilibrium. So, if this
situation is met we shall admit that some assumption is wrong. It might be
argued that the lack of solution can be due to the point-mass approximation
and/or the truncation at third order in $l/s$, whose validity become
increasingly deficient for small values of $s$. However, as pointed out,
these approximations are reasonably good even in the most unfavorable case of
systems in physical contact. The configuration ($M'$, $s$) being what it is,
the only wrong assumption must concern the assumed shape of the density
profile. Indeed, for too strong a tidal interaction, the perturbed system
would be tidally truncated or would not have accreted the surplus material,
accomodating its structure to the nearest available state of equilibrium.
Hence, for consistency with the overall scheme, in this ``saturate regime''
we shall simply take the minimum allowed axial ratio of equilibrium as the
effective solution.

\sec{3. DISTRIBUTION OF PROJECTED AXIAL RATIOS}

The morphological analysis of rich relaxed clusters shows that their
``circularized'' galaxy number density profiles are consistent with a
universal radial profile of the modified Hubble law form, with core radius
$r_0$ equal to $0.25\;h^{-1}$ Mpc (Dressler 1978; Binggeli 1982;
Salvador-Sol\'e et al. 1992), and index $\alpha^g$ equal to 3/2 (Dressler 1978;
West, Dekel, \& Oemler 1987; Salvador-Sol\'e et al. 1992). This universal
profile refers, however, just to the central parts of clusters (up to about
one Abell radius) while we are interested here in the density run up to the
edge of the systems. The statistical analysis by means of the cluster-galaxy
cross-correlation function, of the angular distribution of galaxies around
the position of rich Abell clusters (Lilje \& Efstathiou 1988) shows that
this is consistent with a function of the form (see also Peebles 1980)
$$
\xi_{cg}(r)\;=\;\biggl({hr\over 6}\biggr)^{-2.3}+\biggl({hr\over
7}\biggr)^{-1.7}
\eqno(20)
$$
(with $r$ in Megaparsecs) for $1$ Mpc$\le hr\le 20$ Mpc. According to the
statistical meaning of $\xi_{cg}$ expression (20) should be proportional to
the galaxy number density profile in cluster/halos corrected from the mean
cosmological number density as needed. At small radii, expression (20) is
consistent with a corrected galaxy distribution in the relaxed part of
clusters following a modified Hubble law with index $\alpha^g=1.15$, whereas,
at large radii, it is consistent with a power law of logarithmic slope $-1.7$
or, equivalently, with index $\alpha_h^g$ for the galaxy component in the
outer halo about equal to $0.85$. It is important to remark that the shape of
$\xi_{cg}$ at large radii may be influenced by the clustering of galaxy
clusters, the observed behavior being the geometrical mean of both
contributions. However, since this latter contribution is necessarily close
to a power law of logarithmic slope $-1.8$ (Bahcall 1988), if there is a
significant contribution of cluster halos, this must also have the same form.
(The only halo whose morphology has been studied in detail, the Virgo
Supercluster, shows this kind of profile, with $\alpha_h^g \simeq 1$; Yahil
1974; Yahil, Sandage, \& Tammann 1980.) On the other hand, the roughly
uniform l.o.s. velocity dispersion $\sigma_{los}$ of galaxies in clusters
implies a value of the power index $\alpha$ for the total mass density
profile of clusters roughly equal to 1, in agreement with what is expected
from violent relaxation (Lynden-Bell 1967). This is readily seen from the
hydrostatic equation for the galactic component in spherical clusters and the
assumed universality of the equivalent radial profile. To estimate the value
of index $\alpha_h$ for the total mass in the halo we must make some
assumption on the radial run of the $M/L$ ratio there. Since the infalling
material has the same infall velocity irregardless of its nature and well
mixed  material (with uniform $M/L$) seems a reasonable guess for the
initially unperturbed medium around protoclusters, we shall assume that light
traces mass in cluster halos and take $\alpha_h=\alpha_h^g\simeq0.85$.
(Results are, anyhow, quite insensitive to the value of this latter
parameter, as in the case of $R_h$.)

A rough estimate of the value of $R$ is given by the velocity dispersion of
galaxies times the age of the universe. Such a radius probably embraces the
region of rebounding although not yet fully virialized layers. However, we do
not need matter to be strictly virialized but only with negligible streaming
velocity. Poor clusters show a clear correlation between $\sigma_{los}$ and
Abell (1958) richness $N$, while rich clusters show a much larger scatter
around a roughly fixed value of $\sigma_{los}$ (Bahcall 1981). This behavior
leads to $R \simeq 0.14\, N\,h^{-1}$ Mpc (for $\Omega = 1$) for $N\leq 50$,
and $R \simeq 7\;h^{-1}$ Mpc for $N\geq 50$. (Note that these values are
fully consistent with the radius at which there is the change between the two
extreme power-law regimes of $\xi_{cg}$, this latter function having been
obtained from rich clusters.) A similar situation probably holds for $R_h$.
According to the observed cluster-galaxy cross-correlation this latter radius
would be, for rich clusters, of at least $20\;h^{-1}$ Mpc and possibly as
large as $\sim 40\; h^{-1}$ Mpc (Seldner and Peebles 1977; Peebles 1980),
while for poor and moderately rich clusters it is certainly much smaller. For
this reason we shall simply take $R_h=30\;(N/50)\;h^{-1}$ Mpc for $N\leq 50$,
and $R_h=30\;h^{-1}$ Mpc for $N\ge 50$, in the case that the perturbing
system is more distant than twice the corresponding value of $R_h$, and half
the separation $s$ for closer systems. (Results are also quite insensitive to
this upper bound value of $R_h$.) Note that, at separations smaller than
twice the theoretical value of $R$, the total radius $R_h=s/2$ will get
smaller than $R$, so not only will there be no halo, but the relaxed part of
clusters will then be truncated at $R_h$.

The dependence of $Q$ on cluster richness can be  inferred from the
generalized version of the virial theorem (\S\ 2) for the galactic
component of clusters. By dividing it by the mass $M^g$ we obtain
$$
3\,\sigma_{los}^2\;=\;G\;\rho(0)\;R^2\;f_I^g\;\widetilde
f_U^g\;J(\beta,\beta^g),
\eqno(21)
$$
{}From the definitions of $f_I^g$ and $\widetilde f_U^g$ (eq. \lbrack
14$b$\rbrack\ and \lbrack 16$a$\rbrack\ for the galactic component,
respectively) we have that factor $\rho(0)\;R^2\;f_I^g\;\widetilde f_U^g$ is
proportional to the total mass $M$ over some characteristic length of the
cluster. So one recovers the usual expression of the virial theorem for
isolated spherical systems except for the function $J(\beta,\beta^g)$. The
product $f_I^g\;\widetilde f_U^g$ is essentially proportional to
$R^{-2\alpha}$. Thus, given the proportionality between $R$ and
$\sigma_{los}$, equation (21) leads to the fact that $\rho(0)\,R^{-2\alpha}\,
J(\beta,\beta^g)$ should not depend on cluster richness. Since the function
$J(\beta,\beta^g)$ will typically take the value corresponding to the
saturate regime (see below) it will not depend on cluster richness neither.
So the central density $\rho(0)$ must be proportional to $N^{2\alpha}\sim
N^2$ or, equivalently, $Q$ must be proportional to the square of the richness
ratio. (This dependence of $\rho(0)$ with $N$ leads to a $M/L$ ratio, {\it up
to a fixed radius}, essentially proportional to $N$.) Notice that, since the
richer the cluster, the harder it is that its axial ratio can systematically
reach the saturate value, equation (21) implies that, for rich clusters, the
empirical correlation $\sigma_{los}$ vs $N$ must show a larger scatter than
for poorer ones. Besides, since $J(\beta,\beta^g)$ will then begin to depend,
on the average, on cluster richness as $N^{\gamma}$ with $\gamma<0$, the
correlation $\sigma_{los}$ vs $N$ must become flatter (disregarding of the
actual dependence of $R$ on $N$ for these clusters). Both trends are in full
agreement with observation.

With these values and/or functionalities of the parameters entering in the
model let us now derive the predicted distribution of tidally-induced
elongations. For a cluster of given richness $N$ in the range for which the
empirical distribution has been obtained (\ie $N\ge 30$), the probability
$P(N,\beta)\;d\beta$ of it having intrinsic axial ratio between $\beta$ and
$\beta + d\beta$ due to the tidal action of neighboring clusters {\it with
richness above some threshold} $N_t$ is
$$
P(N,\beta)\;d\beta\;=\;P_{neig}(N,\beta)\;d\beta\;\overline
P_{neig}(N,\le\beta),
\eqno(22)
$$
where $P_{neig}(N,\beta)\;d\beta$ is the probability of finding one of such
neighbors able to produce the wanted axial ratio, and $\overline
P_{neig}(N,\le\beta)$ is the probability that its tidal interaction be
dominant, which coincides with the probability that there is no other
neighbor of the same kind able to yield a smaller axial ratio.

Probability $P_{neig}(N,\beta)\;d\beta$ for $\beta >\beta_{min}$ is
simply given by
$$
P_{neig}(N,\beta)\;d\beta\;=\;4\pi\;n_c \int_0^\infty \;\bigl(1 +
\xi_{cc}(s)\bigr)\;
s^2\; {\cal N}_c(N_{neig}(N,\beta,s)\bigr)\; \biggl|{\partial
N_{neig}(N,\beta,s)\over \partial \beta}\biggr|\; d\beta\;ds.
\eqno(23)
$$
In equation (23) $n_c$ and $\xi_{cc}$ are the mean number density and
correlation function of clusters with $N\ge N_t$, ${\cal N}_c(N)$ is the
normalized cluster richness function, and $N_{neig}(N,\beta,s)$ is the
richness necessary for a neighbor at $s$ be able to yield the wanted axial
ratio. This latter value is obtained from equation (17) where, for
simplicity, radii $R$ and $R_h$ are taken equal to those corresponding to a
cluster of median richness $N_{med}$ in the allowed range for perturbers.
Note also that in equation (23) we have neglected any spatial segregation
among clusters with different richnesses within this range. These two
practical approximations have noticeable consequences discussed below. The
cluster richness function ${\cal N}_c(N)$ is approximated by two power laws
of indexes $-2$ at the poor end and $-5$ at the rich one, matching at
$N^*\simeq 65$ (Bahcall 1979). This allows us to calculate the value
$N_{med}$ and the mean density $n_c$ for any threshold $N_t$ from some given
value of reference; here we use $n_c=1.0\;10^{-5}\;h^3$ clusters per
Megaparsec for $N_t=50$ (Bahcall 1988). Finally, the correlation function
$\xi_{cc}(s)$ for threshold $N_t$ is taken equal to $A\,s^{-1.8}$ for $s\le
50\;h^{-1}$ Mpc and null beyond that distance, with the correlation amplitude
$A$ depending on cluster richness according to the empirical relation
$A\propto N_{med}$ (Bahcall \& West 1992, and references therein). Notice
that factor $d\beta$ in the right hand side of equation (23) allows us to
calculate the probability of finding one neighbor with the appropriate
richness by simply taking the integral over the volume of the density
probability of finding it. Although the integral extends to the whole space,
one can take a finite though sufficiently large piece of universe warranting
the convergence of the result. It may also be more realistic to exclude some
{\it small} volume around the perturbed system, but this makes no appreciable
difference in the final results.

On the other hand, probability $\overline P_{neig}(N,\le\beta)$ is given by
$$
\overline P_{neig}(N,\le\beta)\;=\exp\Biggl[- \;4\pi\;n_c \int_0^\infty
\;\bigl(1 + \xi_{cc}(s)\bigr)\; s^2\; {\cal N}_c\bigl(\ge
N_{neig}(N,\beta,s)\bigr)\;ds\Biggr],
\eqno(24)
$$
with the exponent of the right-hand-side member equal to minus the expected
number of neighbors able to yield an axial ratio smaller than or equal to
$\beta$, which coincides with the number of neighbors with richness larger
than or equal to that yielding the wanted axial ratio. Taking into account
that expression (23) is equal to the partial derivative of the exponent in
expression (24), probability $P(N,\beta)\;d\beta$ can  be written as minus
the partial derivative with respect to $\beta$ of $\overline
P_{neig}(N,\le\beta)$ or, equivalently,
$$
P(N,\ge\beta)\;=\;\overline P_{neig}(N,\le\beta).
\eqno(25)
$$
{}From equation (25) it can be readily seen that the probability
$P(N,\beta)\;d\beta$ is correctly normalized to unity for each richness $N$
of the perturbed system.

The distribution of intrinsic axial ratios $\Psi(\beta)$ for $\beta
>\beta_{min}$ of clusters with $N\ge 30$ can be thus obtained by deriving
with respect to $\beta$ the integral over $N$ of probability $P(N,\ge
\beta)$, given by equation (25), weighted by ${\cal N}_c(N)$. And the value
of the distribution function at $\beta =\beta_{min}$ can be readily
calculated by taking into account that the whole integral of $\Psi$ over
$\beta$ must be unity. This leads to the following expression
$$
\Psi(\beta)\;=\;{\partial\over\partial \beta} \int_{30}^\infty \;\overline
P_{neig}(N,\le\beta)\;{\cal N}_c(N)\;dN
$$
$$
\;+\;\delta(\beta - \beta_{min})\;\biggl(1-\int_{30}^\infty \;\overline
P_{neig}(N,\le\beta_{min})\; {\cal N}_c(N)\;dN\biggr).
\eqno(26)
$$

Finally, in order to obtain the distribution of axial ratios for the galactic
component we must simply take into account that each axial ratio $\beta$
gives rise to an ``observable'' axial ratio $\beta^g$, related to the former
through equation (19). So the wanted distribution of axial ratios
$\Psi(\beta^g)$ is
$$
\Psi(\beta^g)\;=\;\Psi(\beta)\;{d \beta\over d\beta^g}
\eqno(27)
$$
in terms of the distribution $\Psi(\beta)$ given by equation (26) and the
function $\beta(\beta^g)$ given by equation (19) (with $R$ and $R_h$ in
$\widetilde f_U$ and $\widetilde f_U^g$ taking the values for perturbers with
richness equal to $N_{med}$ as above). Notice that the distribution
$\Psi(\beta^g)$ also has a Dirac delta as high as that of $\Psi(\beta)$ but
shifted to $\beta_{min}^g=\beta^g(\beta_{min})$. The resulting distribution
of intrinsic axial ratios $\beta^g$  (for a richness threshold of perturbers
equal to 45; see discussion below) is plotted in Figure 1a. The solution
shows a very sharp peak at $\beta^g_{min}\simeq 0.5$ in agreement with what
is required by observation (Plionis et al. 1991). However, it notably differs
from a Gaussian function centered on that value and standard deviation equal
to $0.15$. The desagreement seems to be twofold: the peak at $\beta^g_{min}$
is too sharp, and there is an important bump at large axial ratios. But
before drawing any definite conclusion we should first look at the
distribution of projected axial ratios.

By using the relation between the distributions of intrinsic and projected
axial ratios, $\beta$ (or $\beta^g$) and $\beta_p$ ($\beta_p^g$) for prolate
spheroids (Hubble 1926)
$$
\Phi(\beta_p)\;=\;{1\over\beta_p^2}\;
\int_0^{\beta_p}\;{\beta^2\;\Psi(\beta)\; d\beta
      \over (1-\beta^2)^{1/2}\;(\beta_p^2-\beta^2)^{1/2}}.
\eqno(28)
$$
we can obtain the really compelling distribution, $\Phi(\beta_p^g)$, from the
intrinsic one, $\Psi(\beta^g)$, given by expression (27) and directly compare
it with observation. In Figure 1b we plot the function $\Phi(\beta_p^g)$
corresponding to the distribution in 3D shown in Figure 1a, together with the
empirical distribution obtained by Plionis et al. (1991). In fact to properly
compare both distributions we must take into account measuring errors.
According to Plionis et al. (1991) the standard deviation in the axial ratio
obtained by his method, calculated from Monte Carlo clusters, is equal to
$0.05$--$0.07$, independently of the specific value of the intrinsic axial
ratio simulated. The scatter shown by their data on real clusters is somewhat
larger, with average standard deviation equal to $0.10$ (see the data quoted
by Plionis et al. 1991 for different thresholds of cluster density), such an
increase being probably caused by the varying center location and background
contamination which are both absent in the simulations. Thus, in order to
mimic the effects of measuring errors we have convolved our theoretical
distribution by a Gaussian function with standard deviation equal to $0.10$.
As shown in Figure 1c, the maximum at $\sim 0.6$ of the convolved theoretical
distribution of projected axial ratios is now sufficiently smooth. So the
sharpness of the Dirac delta at 3D is actually not any problem. Projection
and convolution has also somewhat mitigated the bump at large axial ratios,
but there is still a very marked secondary maximum at $\beta_p\sim 0.90$,
fully absent in the empirical histogram. This would only disappear for a high
enough Dirac delta in 3D.

In diminishing the threshold allowed for perturbers one should in principle
obtain more marked elongations and, consequently, a higher Dirac delta.
Indeed, all rich perturbers causing the previous elongations are always
included, and one is just adding new perturbers, which should tend to
increase the resulting elongations. In practice, however, such an expected
trend is only followed at large thresholds; at small ones the solution shows
just the opposite behavior. This can be seen in Figure 2 where we have
plotted the height of the Dirac delta at $\beta^g_{min}$ of the distribution
in 3D as a function of the threshold $N_t$. The wrong behavior shown by the
solution at small thresholds is just a consequence of the use of average
characteristics for clusters above some given threshold (see comments on eq.
\lbrack 22\rbrack). In lowering the threshold we are taking smaller $R$ and
$R_h$ and a smaller correlation amplitude. The new values are well suited
{\it on the average} for clusters in the new range of richnesses, but worse
suited for the richests ones in this range, which rather partake of the
preceding average characteristics. This introduces an error which gets more
and more severe for increasingly lower thresholds. Only for large enough
thresholds (above $N^*= 65$) can the effect not balance the dramatic increase
in the number of perturbers when diminishing it, so that the good behavior of
the distribution function there is guaranteed. From this discussion it is
clear that: (1) the lower the threshold, the more underestimated the effects
of tidal interaction in the derived distributions of axial ratios, (2)
correction of this flaw should lead to, at least, the distribution with the
most marked axial ratios obtained, (3) very rich clusters are not numerous
enough and, hence, they are placed typically at too large distances from the
perturbed system to systematically yield very marked elongations, and (4)
poor clusters are, on the contrary, quite numerous and can be found nearer to
the perturbed system, but they are not rich enough to yield very marked
elongations neither. From Figure 2 we see that the most marked elongations
correspond to a threshold $N_t$ of about 45; this was the value used in
Figure 1. Note that, since the values of $R$ and $R_h$ keep fixed for $N_t\ge
35$ ($N_{med}=50$), the distribution plotted in that figure is only affected
by the neglect of the segregation in cluster richness, which is not a very
important effect.

There is, however, another simplifying assumption in the model that may have
important consequences in the final result. In the derivation above we have
neglected the tidal action of non-dominant neighbors. The reason for this is
that the total number of neighbors located within the distance necessary for
their tidal action be relevant (see \S\ 4), is relatively small (about 14
with $N\ge 45$ and $s\le 55\;h^{-1}$ Mpc). This guarantees that, in any
actual realization, there are important gaps in the distributions of neighbor
richnesses and separations. Consequently, the tidal force of any non-dominant
neighbor will usually be negligible compared to that of the dominant one.
However, clusters tend to be clustered and non-dominant neighbors will be
preferentially located near to the dominant one, boosting its tidal action.
Besides, we have also neglected the tidal effects of voids (Ftaclas 1983)
which should also boost the tidal action of groups of clusters. So the
previous simple approach has been rather underestimating the true
tidally-induced elongation of clusters. An accurate prediction of the
distribution of cluster axial ratios through the vectorial composition of the
tidal force caused by each individual neighbor and void is hopeless because
this demands a full statistical knowledge of the spatial distribution of
clusters, while only the very first $N$-point correlation functions are
available. Nonetheless, it would be interesting to have an idea of the
effects of clustering. In fact, a straighforward extension of the previous
model can be developed which accounts very approximately for the tidal action
of groups of clusters, or superclusters, and voids amidst them making only
use of the cluster two- and three-point correlation functions.

Taking advantage of the point mass approximation, we can consider all
neighboring clusters within spheres centered on one of them and with radius
equal to the distance of this center to the perturbed cluster as contributing
with the sum of their masses to the potential of the central perturbing
system. One must correct, of course, this composite mass from the
contribution inside the sphere of the mean uniform density of clusters. So we
are only concerned with the number of clusters ``in excess'' inside these
spheres. (Note that we should strictly add, on the contrary, the previously
substracted contribution of the mean density in individual clusters, but this
is a small correction that can be neglected.) This procedure would accurately
account for the tidal effects of any expected anisotropy in the distribution
of neighboring clusters around a given one if the density distribution around
the center of these spheres were really a continuous spherically symmetric
function. Actually, the probability of finding clusters inside the spheres
does not only depend on the radial distance from their center but also, to a
smaller extent, on the distance to the perturbed cluster. (For simplicity
this latter dependence is neglected in the calculations; taking it into
account should result in slightly more marked elongations.) Besides, the
number of clusters in excess contained in these spheres is not very large. So
the associated mass density distribution will notably deviate from a
continuous function. (We take it as simply proportional to the probability of
finding clusters). However, we are interested in {\it estimating} the effects
of anisotropies in the spatial distribution of clusters {\it in a statistical
manner} rather than trying to compute the accurate tidal force exerced on a
cluster in any given realization. So our approximate procedure is justified
and should actually lead to quite a good estimate of the real distribution of
axial ratios.

In Appendix A we derive the equations leading to the distribution of axial
ratios according to this much more accurate version of the model. Because of
the effects mentioned above, we have tried different thresholds, the most
marked elongations being now obtained for $N_t\simeq 35$. The solution is
shown in Figure 3. (As in the preceding version the maximum length of
positive correlation has been taken equal to $50\;h^{-1}$ Mpc. It is
worthwhile mentioning that in the present version, the larger this length,
the more marked the resulting elongations. In fact, for values greater than
$\sim 60 h^{-1}$ Mpc they would even be excessively marked, but this effect
should not be taken too seriously because it might be balanced by a cluster
richness function steeper than assumed at very large richnesses.) For
comparison with Figure 1 we plot the distribution of intrinsic and projected
axial ratios, the latter in the real as well as degraded versions. As
expected, the inclusion of the tidal action of groups of clusters
(superclusters) has notably increased the height of the Dirac delta at
$\beta_{min}$ in the distribution in 3D and erased the second maximum at
about 0.9 in the distribution in 2D. The result is in very good agreement
with observation, which is especially remarkable given the simplifying
assumptions involved in the model and the fact that it has no free
parameters. (Note that even the small inflexion observed at an axial ratio of
0.9 in the predicted distribution and absent in the empirical one woud tend
to disappear if we could accurately correct the effects of measuring errors
at the upper bound value of one. Indeed, the convolution by a gaussian
function introduces a small edge effect which tends to slightly
underestimates the true elongations there.)

\sec{4. DISCUSSION}

Dominant perturbers are, therefore, typically rich ($N\ge 45$) single
clusters and groups of clusters with a slightly wider range of richnesses
($N\ge 35$). In any case, the nearest neighboring cluster with $N\ge 30$ is
typically not the dominant perturbing system, which explains the lack of
significant correlation between cluster orientation and angular position of
the nearest neighboring cluster. Our results also explain the existence, on
the contrary, of a correlation between cluster orientation and angular
position of {\it any} neighbor within some given distance. The predicted
value of this distance can be easily obtained from the theoretical
distribution of axial ratios by imposing increasingly large minimum
separations to the possible perturbers. In doing so, the typical elongation
of clusters, measured from the height of the Dirac delta at $\beta_{min}^g$
in 3D, gets smaller and smaller, and, for a large enough minimum separation,
it becomes so small that no significant alignment would be found between the
perturbed cluster and the perturbing ones. This upper value of the minimum
separation determines, therefore, the maximum distance for significant
alignments. By taking the fraction of typically elongated clusters equal to
10 \% (5 \%)  we are led to a value of that distance equal to 55 $h^{-1}$ Mpc
(75 $h^{-1}$ Mpc), in very good agreement with observation, too. (The
predicted values may seem slightly larger than observed, but this is not
important because by using the height of the maximum at 2D instead of the
height of the Dirac delta at 3D we would have obtained smaller values.)

The same tidal action causing the elongation of relaxed clusters should also
tend to elongate individual galaxies. Because of their much higher
concentration, the resulting elongation should be, however, much less marked
than for clusters (eq. \lbrack 19\rbrack). In other words, the shape of
galaxies would not be supported, in general, by the tidal action of clusters.
However, this might easily cause an appreciable elongation to giant D and cD
galaxies. So the proposed scenario would also explain the observed alignment
between clusters and their first ranked galaxies. Note that, for the same
reason, some alignment should also be found between clusters and any stable
substructure inside them such as binary systems (if any). This might explain
the recent finding by Trevese, Cirimele, \& Flin (1992). In summary, the
proposed scenario is in agreement or, at least, consistent with all reported
alignment effects involving clusters.

The model of tidal interaction developed here assumes the velocity tensor of
galaxies in clusters isotropic. As pointed out, some anisotropy is forseeable
(although difficult to characterize) coming from the aspherical configuration
of the protocluster. Since this would also be tidally-induced by the same
dominant perturber as, later on, once the cluster has virialized, such an
anisotropy should not prevent the cluster or, more exactly, its anisotropic
galactic component to reach the very marked elongation found under the
isotropic condition. On the contrary, it should rather favor a more marked
elongation. Therefore, the axial ratio in the saturate regime of tidal
interaction for such anisotropic systems should not be too different from
that found in the isotropic aproximation. Otherwise, the predicted typical
tidally-induced elongation of clusters might turn out to be unacceptably
high. This should be possible to check by means of N-body simulations.

Since the isotropic condition is very well suited for the intracluster
medium, the model developed here could be readily applied, without any
limitation of the previous kind, to this gaseous component. Unfortunately,
there is so far no well-established distribution of projected axial ratios of
clusters drawn from their X-ray images. Work in progress in this line seems
to point at substantially less marked typical elongations (Jones \& Forman
1991). However, this might simply be due to the fact that clusters studied in
X-rays tend to be very rich. So it would be of major interest to see whether
or not the observed elongation of the gaseous component of clusters is also
in agreement with that predicted by the present model for the appropriate
range of cluster richnesses.

As an important byproduct of the proposed scenario we have that tidal
interaction between clusters is typically in the saturate regime, which
explains in a very natural manner the ``universal'' value of their intrinsic
axial ratio. Therefore, the growth and dynamical state of clusters should be
notably influenced by this interaction, and in neglecting it one might be
committing non-negligible systematic errors. This is apparently the case for
the usual estimate of cluster masses by means of the virial theorem in its
version for spherically symmetric isolated systems.

\sec{Aknowledgements}

We thank P.J.E. Peebles for a fruitful discussion, and M. Castillo and J.
Carvalho for their aid in a preliminary study. This work has been supported
by the Direcci¢n General de Investigaci¢n Cient\'\i fica y T\'ecnica, under
contract PB89-0246, and PB90-0448.

\page
\sec{APPENDIX A}
\sec{Multiple Tidal Interaction}

The probability $P(N,\beta)\,d\beta$ for a cluster of given richness $N$
to have axial ratio between $\beta$ and $\beta + d\beta$ is equal to the
sum of probabilities of this axial ratio being caused by {\it
one single} neighbor in excess relative to the mean cluster density,
$P_1(N,\beta)\;d\beta$, by a group of {\it two} neighbors in excess,
$P_2(N,\beta)\;d\beta$, by a group of {\it three} neighbors in excess,
$P_3(N,\beta)\;d\beta$, etc...
This multiplicity of galaxies in excess refers to inside
spheres $\delta V$ centered on one neighboring cluster (which guarantees
some net excess of clusters in the sphere)
and with radius equal to $s$.

The probability $P_1(N,\beta)\;d\beta$ is given by equation (22), but for the
fact that the probability $P(N,\beta)$ includes the condition that the
perturbing neighbor be really isolated in the sphere once it has been
corrected from the mean cosmological density of clusters, and the probability
$\overline P_{neig}(N,\le\beta)$ is replaced by the probability that there is
no other {\it isolated nor grouped} neighbors in excess able to produce a
smaller value of $\beta$. On the other hand, the probability
$P_2(N,\beta)\;d\beta$ is given by an equation similar to (22) with the
product in the right-hand-side member involving the probability of finding a
group of two systems in excess able to jointly produce the wanted axial ratio
or, equivalently, the corresponding additive value of $G(\beta)$, and the
probability of this tidal action be dominant, which coincides with the
corresponding probability for the isolated case; we shall write this
probability as $\overline P(N,\le \beta)$. And so on. Therefore, we have
$$
P(N,\beta)\;d\beta\;=\;\Big[P_1(N,\beta)+P_2(N,\beta)+...\Big]\;\overline
    P(N,\le \beta)\;d\beta.
\eqno({\rm A}1)
$$

The probability $\overline P(N,\le \beta)$ is given by
$$
\overline P(N,\le \beta)\;=exp\Biggl\{- \,4\pi\,n_c \int_0^\infty
ds\,s^2\;\bigl(1 + \xi_{cc}(s)\bigr)\;e^{-\nu(s)}\;
$$
$$
\times\biggl[\,{\cal N}_1(N,\ge G(\beta),s)) + {1\over 2}\,\nu(s)\,{\cal
N}_2(N,\ge G(\beta),s)+ {1\over 3}\,{\nu^2(s)\over 2!}\,{\cal N}_3(N,\ge
G(\beta),s)+... \biggr]\Biggr\}, \eqno({\rm A}2)
$$
with
$$
\nu(s)\;=\;n_c\int\limits_{\delta V}
\;{\xi_{cc}(s)+\xi_{cc}(r)+\xi^2_{cc}(s)+2\,\xi_{cc}(s)\,\xi_{cc}(r)\over
1+\xi_{cc}(s)}\;dV \eqno({\rm A}3)
$$
giving the mean number of neighbors in excess, apart from the central one,
within a sphere $\delta V$ centered on a galaxy at $s$ from the perturbed
one. In equation (A3) we have taken the reduced three-point correlation
function in the simple form of the {\it Kirkwood superposition} consistent
with observation (Jing \& Valdarnini 1991 and references therein), and we
have approximated the two-point correlation of all neighbors in each sphere
with respect to the perturbed cluster by that of the central neighbor
(warranting in this way the necessary spherical symmetry of the density
distribution inside the spheres). Note that the different terms in the
exponent of equation (A2) contain a factor (\eg, ${1\over 2}$, ${1\over3}$,
etc...) which corrects from repeating the configurations when integrating
over the spatial location of the central neighbor. Functions ${\cal
N}_i(N,\ge G(\beta),s))$ in equation (A2) give the probability that a group
of $i$ neighbors located at $s$ be able to yield an axial ratio smaller than
or equal to $\beta$ (or, equivalently, a value of the additive quantity $G$
greater than or equal to $G(\beta)$). In terms of the cluster richness
function they write
$$
{\cal N}_1(N,\ge G(\beta),s)\;=\;{\cal N}_c(\ge N_{neig}(N,\beta,s)),
\eqno({\rm A}4a)
$$
and the iterative relation for $i\ge 2$
$$
{\cal N}_i(N,\ge G(\beta),s)\;=\;
{\cal N}_c(\ge N_t)\;{\cal N}_{i-1}\bigr(N,\ge (G(\beta)-G_t),s\bigr)
$$
$$
\int^{\sqrt{N_{neig}^2(N,\beta,s)-N^2_t}}_{N_t} \;{\cal N}_c(\widetilde
N)\;{\cal N}_{i-1}\bigl(N,\ge (G(\beta)-\widetilde G),s\bigr)\;
d\widetilde N \eqno({\rm A}4b)
$$
(with $G_t$ and $\widetilde G$ standing for the values of $G$ yielded by a
neighbor of richnesses $N_t$ and $\widetilde N$, respectively) for
$N_{neig}(N,\beta,s)\ge \sqrt 2\, N_t$, and ${\cal N}_i(N,\ge G(\beta),s)=1$
for $N_{neig}(N,\beta,s)> \sqrt 2\, N_t$ (notice that in relation \lbrack
A4$b$\rbrack\ ${\cal N}_c(\ge N_t)$  is actually equal to one). Given that
these expressions make it hard to calculate the exponent of the
right-hand-side of equation (A2) it is better to take, for $i\ge 3$, some
approximate but more practical expressions. The very steep cluster richness
function allows us to write
$$
\int^{\sqrt{N_{neig}^2(N,G(\beta),s)-N^2_t}}_{N_t} \;{\cal N}_c(\widetilde
N)\;{\cal N}_{i-1}\bigl(N,\ge (G(\beta)-\widetilde G),s\bigr) \;
d\widetilde N\;\simeq $$ $$
\;{\cal N}_{i-1}\bigl(N,\ge G(\beta)-G_t,s\bigr)\;\Bigl[{\cal N}_c(\ge
N_t)-{\cal N}_{i-1}\bigl(N,\ge
(G(\beta)-G_t),s\bigr)\Bigr], \eqno({\rm A}5)
$$
so that the iterative relation (A4$b$) takes the simple form
$$
{\cal N}_i(N,\ge G(\beta),s)\;\simeq\; {\cal N}_1\bigl(N,\ge
(G(\beta)-G_t),s\bigr)\; \biggl[2\;-\;{\cal N}_1\bigl(N,\ge
(G(\beta)-G_t),s\bigr)\biggr]^{i-1}.
\eqno({\rm A}6)
$$
(Note that for $N_{neig}(N,\beta,s)> \sqrt 2\, N_t$ one has $G(\beta)-G_t\le
G_t$, and functions ${\cal N}_i(N,\ge G(\beta),s)$ given by equation \lbrack
A6\rbrack\ reduce to unity as required.) Finally, by substituting functions
${\cal N}_i(N,\ge G(\beta),s)$ given by the approximate relations (A6) into
equation (A2) we are led to
$$
\overline P(N,\le \beta)\;=\;exp\Biggl\{- \,4\pi\,n_c \int_0^\infty
ds\,s^2\;\bigl(1 + \xi_{cc}(s)\bigr)\,e^{-\nu(s)}\, \biggl[\,{\cal N}_1(N,\ge
G(\beta),s)
$$
$$
-{\cal N}_1\bigl(N,\ge (G(\beta)-G_t),s\bigr)+{{\cal N}_1\bigl(N,\ge
(G(\beta)-G_t),s\bigr) \over \nu(s)\bigl[2-{\cal N}_1\bigl(N,\ge
(G(\beta)-G_t),s\bigr)\bigr]}
$$
$$
\times\biggl(\,e^{\nu(s)\bigl[2-{\cal N}_1\bigl(N,\ge
(G(\beta)-G_t),s\bigr)\bigr]}\;-\;1\,\biggr)\biggr]\Biggr\}.
\eqno({\rm A}7)
$$
It is worthwhile pointing out that approximation (A5) slightly
underestimates the elongations of the exact solution.

Like in the one-single-neighbor version (see \S\ 3), the sum
$P_1(N,\beta)\,+\,P_2(N,\beta)\,+...$ in equation (A1) is equal to the
partial derivative with respect to $\beta$ of the exponent in the
right-hand-side member of equation (A2). So we always have the practical
relation
$$
P(N,\ge\beta)\;=\;\overline P(N,\le\beta),
\eqno({\rm A}8)
$$
whose integration with respect to $N$, appropriately weighted by ${\cal
N}_c(N)$, and subsequent derivation with respect to $\beta$ readily
leads to the wanted distribution of intrinsic axial ratios.

The volume of spheres $\delta V$ increases indefinitely with increasing $s$.
But the number of neighbors in excess from the uniform mean density inside
them stabilizes when a distance is reached for which the radius of the
corresponding sphere becomes larger than the maximum length of positive
correlation of clusters, \ie the typical size of superclusters. Thus, the
probability of finding a group of $i$ neighbors in excess within these
spheres with appropriate richnesses to yield the wanted axial ratio will
diminish very rapidly beyond that distance, which limits the distance of
significant tidal interaction (see \S\ 4).

\page
\sec{APPENDIX B}
\sec{Potential Energy of the Galactic Component}

The potential energy tensor of a self-gravitating system is equal to
$$
U_{ij}\;=\;-{1\over 2}\,G\;\int \int \rho({\bf x})\;\rho({\bf
x}')\;{(x_i-x_i')\;(x_j-x_j')\over |{\bf x}-{\bf x}'|}\; d{\bf x}\;d{\bf x}'.
\eqno({\rm B}1)
$$
Taking advantage of factor 1/2, in the case of self-similar ellipsoids (\ie
$\rho=\rho(m)$, with $m^2={x_1^2\over a_1^2}+{x_2^2\over a_2^2}+{x_3^2\over
a_3^2}$) equation (B1) can be written in the practical form (Roberts 1962)
$$
U_{ij}\,=\;-\;\int\int\limits_{\kern -1.2 em m\ge
m'}\;\rho(m')\;\Phi_{ij}(m)\;d{\bf m}\;d{\bf m}', \eqno({\rm B}2)
$$
with $\Phi_{ij}(m)$ the tensor potential produced by the homoeoid ($m$,
$m+dm$). But the tensor potential of an homoeoid is constant inside it, so
that expression (B2) allows one to readily calculate the potential energy
tensor, the result being equation (13) (Roberts 1962). This development is,
of course, only possible if the component for which we want to calculate the
potential energy tensor is {\it self-gravitating} (otherwise there would not
be factor 1/2). Since the density distribution of the galactic component of a
galaxy cluster $\rho^g$ does not cause by itself the whole tensor potential
$\Phi_{ij}$ but only contributes marginally to it, we cannot apply this
method.

To calculate the potential energy tensor of the galactic component let us
decompose the total density $\rho$ into the sum of two densities, that of
galaxies, $\rho^g$, and the remaining one $\rho^r=\rho - \rho^g$. Then the
integral on the right-hand-side of equation (B1) decomposes in a sum of three
integrals involving, respectively, half the product $\rho^r\,\rho^r$, half
the product $\rho^g\,\rho^g$, and product $\rho^r\,\rho^g$. The sum of the
latter two integrals exactly gives the wanted potential energy tensor
$U_{ij}^g$. Therefore, this can be calculated as the total potential energy
tensor $U_{ij}$ given by equation (B1) minus the potential energy tensor
$U^r_{ij}$ associated to a {\it self-gravitating} fluid with density $\rho^r$
for which the method above also applies. By doing so we are led to
$$
U^g_{ij}\;=\; -2\pi\; G\; \rho(0)\; A_i(\beta)\; I^g_{ij} \; f^g_U,
\eqno({\rm B}3)
$$
where we have approximated the axial ratio of component $r$ to that of the
whole mass distribution (this is a very good approximation because the shape
of the density profile of the former is essentially equal to that of the
whole mass), with $I^g_{ij}$ defined as $I_{ij}$ (eq. \lbrack 12\rbrack) in
terms of the density profile $\rho^g$, and $f^g_U$ given by
$$
f^g_U\;=\;{f_I\,f_U\;-\;f^r_I\,f^r_U\over f_I^g}.
\eqno({\rm B}4)
$$
For simplicity, the present proof has been carried out for the potential
energy tensor of the galactic component in the whole system and not in a
limited part of it. The changes that should be introduced in this latter case
are given in \S\ 3.

\page
\refhead

\ref{Aarseth, S.J., \& Binney, J. 1978, MNRAS, 185, 227}
\ref{Abell, G.O. 1958, ApJS, 3, 211}
\ref{Argyres, P.C., Groth, E.J., Peebles, P.J.E.,\& Struble, M.F. 1986, AJ, 91,
471}
\ref{Bahcall, N.A. 1979, ApJ, 232, 689}
\ref{Bahcall, N.A. 1981, ApJ, 247, 787}
\ref{Bahcall, N.A. 1988, ARAA, 26, 631}
\ref{Bahcall, N.A., \& West, M.J. 1992, ApJ, 392, 419}
\ref{Bardeen, J.M., Bond, J.R., Kaiser, N., \& Szalay, A.S. 1986, ApJ, 304, 15}
\ref{Binggeli, B. 1982, A\&A, 107, 338}
\ref{Binney, J., \& Silk, J. 1981, MNRAS, 188, 273}
\ref{Carter, D., \& Metcalfe, N. 1980, MNRAS, 191, 325}
\ref{Chandrasekhar, S., \& Lebovitz, N.R. 1963, ApJ, 137, 1172}
\ref{Chandrasekhar, S. 1969, Ellipsoidal Figures of Equilibrium (New Haven:
Yale University Press)}
\ref{Dekel, A., West, M.J., \& Aarseth, S.J. 1984, ApJ, 279, 353}
\ref{Di Fazio, A., \& Flin, P. 1988, A\&A, 200, 5}
\ref{Dressler, A. 1978, ApJ, 226, 55}
\ref{Dressler, A. 1981, ApJ, 243, 26}
\ref{Flin, P. 1987, MNRAS, 228, 941}
\ref{Fong, R., Stevenson, P.R.F., \& Shanks, T. 1990, MNRAS, 242, 146}
\ref{Frenk, C.S., White, S.D.M., \& Davis, M. 1983, ApJ, 205, 716}
\ref{Ftaclas, C. 1983, MNRAS, 202, 7p}
\ref{Gregory, S.A., \& Tifft, W.G. 1976, ApJ, 205, 716}
\ref{Hubble, E. 1926, ApJ, 64, 321}
\ref{Jing, Y.P., \& Valdarnini, R. 1991, A\&A, 250, 1}
\ref{Jones, C., \& Forman, W. 1992, in The X-ray Background, ed. X. Barcons
and A.C. Fabian, (Cambridge: Cambridge University Press)}
\ref{Lambas, D.G., Groth, E.J., \& Peebles, P.J.E. 1988, AJ, 95, 975}
\ref{Lilje, B., \& Efstathiou, G. 1988, MNRAS, 231, 635}
\ref{Lynden-Bell, D. 1967, MNRAS, 136, 101}
\ref{McGillivray H.T., Martin, R., Pratt, N.M., Reddish, V.C., Seddon, H.,
Alexander, L.W.G., Walker,
G.S., \& Williams, P.R. 1976, MNRAS, 176, 649}
\ref{Oort, J. 1983, ARAA, 21, 373}
\ref{Peacock, J.A., \& Heavens, A.F. 1985, MNRAS, 217, 805}
\ref{Peebles, P.J.E. 1980, in Physical Cosmology,
ed. R. Balian, J. Audouze, \& D.N. Schramm (Amsterdam: North-Holland), p. 213}
\ref{Plionis, M., Barrow, J.D., \& Frenk, C.S. 1991, MNRAS, 249, 662}
\ref{Rhee, G.F.R.N., \& Katgert, P. 1987, A\&A, 183, 217}
\ref{Roberts, P.H. 1962, ApJ, 136, 1108}
\ref{Rood, H.J., Pagel, T.L., Kintner, E.C., \& King, I.R. 1972, ApJ, 175, 627}
\ref{Salvador-Sol\'e, E., Sanrom\`a, M., \& Gonz\'alez-Casado, G. 1992, ApJ, in
press}
\ref{Sastry, G.N. 1968, PASP, 80, 252}
\ref{Seldner, M., \& Peebles, P.J.E. 1977, ApJ, 215, 703}
\ref{Struble, M.F., \& Peebles, P.J.E. 1985, AJ, 90, 592}
\ref{Trevese, D., Cirimele, G., \& Flin, P. 1992, AJ, 104, 935}
\ref{Ulmer, M.P., McMillan, S.L.W., \& Kowalski, M.P. 1989, ApJ, 338, 711}
\ref{West, M.J. 1989a, ApJ, 344, 535}
\ref{West, M.J. 1989b, ApJ, 347, 610}
\ref{West, M.J., Dekel, A., \& Oemler, A. 1987, ApJ, 336, 46}
\ref{Yahil, A. 1984, ApJ, 191, 623}
\ref{Yahil, A., Sandage, A., \& Tammann G.A. 1980, in Physical Cosmology,
ed. R. Balian, J. Audouze, \& D.N. Schramm (Amsterdam: North-Holland), p. 127}

\page

\normal

\sec{Figure Captions}

{\bf Figure 1.} Theoretical distributions of intrinsic (a) and projected (b)
axial ratios predicted by the model of tidal interaction in the simplest
one-single-neighbor version. In panel (c) we plot the same solution as in
panel (b), but conveniently degraded in order to simulate the effects of
measuring errors. The histograms in panels (b) and (c) give the empirical
distribution of projected axial ratios obtained by Plionis et al. (1991).

{\bf Figure 2.} Height of the Dirac delta at $\beta=0.4963$ in the predicted
distribution of intrinsic axial ratios of Figure 1a as a function of the
threshold richness $N_t$ of the perturbing clusters.

{\bf Figure 3.} Same as Figure 1, but for the more accurate version of the
model dealing with the tidal action of single and grouped clusters.

\end